\documentclass[aps,prb,showpacs,twocolumn]{revtex4-1}
\usepackage{amssymb}
\usepackage{amsmath}
\usepackage{graphicx}
\usepackage{dsfont}
\usepackage{times}

\begin{document}

\title{Approaching Thouless Energy and Griffiths Regime in Random Spin
Systems By Singular Value Decomposition}
\author{Wen-Jia Rao}
\email{wjrao@hdu.edu.cn}
\affiliation{School of Science, Hangzhou Dianzi University, Hangzhou 310027, China.}
\date{\today }

\begin{abstract}
We employ singular value decomposition (SVD) to study the eigenvalue spectra
of random spin systems. By SVD, eigenvalue spectrum is decomposed into
orthonormal modes $W_k$ with weight $\lambda_k$. We show that the scree plot
($\lambda_k$ with respect to $k$) in the ergodic phase contains two branches
that both follow power-law $\lambda_k\sim k^{-\alpha}$ but with different
exponents $\alpha$. By evaluating $W_k$, it's verified the part of $%
\lambda_k $ with $k>k_{\text{Th}}$ is universal that follows random matrix
theory, where $k_{\text{Th}}$ is related to the Thouless energy. We further
demonstrate that $\alpha$ corresponds only to the exponential part of the
level spacing distribution while being insensitive to the level repulsion,
or equivalently the system's symmetry. Consequently, $\alpha$ gives an
underestimation for the many-body localization transition point, which
suggests a non-ergodic behavior that may be attributed to the Griffiths
regime.
\end{abstract}

\maketitle

\section{Introduction}

\bigskip The quantum phases of matter in isolated systems is a focus of
modern condensed matter physics, where the existence of two generic phases
has been well-established: an ergodic phase and a many-body localized (MBL)
phase\cite{Gornyi2005,Basko2006}. In an ergodic phase, the system acts as
the heat bath for its subsystem, which results in extensive quantum
entanglement that follows volume-law. In contrast, a MBL phase is where
localization persists in the presence of weak interactions, which leads to
area-law entanglement. The different scaling behaviors of quantum
entanglement provide the modern understanding about these two phases\cite%
{Kjall2014,Yang2015,Serbyn16,Maksym2015,Kim,Bardarson,Abanin,Znidaric,Garcia}%
.

On the other hand, the ergodic and MBL phase are traditionally distinguished
by their eigenvalue statistics\cite%
{Oganesyan,Avishai2002,Regnault16,Regnault162,Huse1,Huse2,Huse3,Luitz,Serbyn}%
, whose foundation is laid by the random matrix theory (RMT)\cite%
{Mehta,Haake}. RMT is a powerful mathematical tool that describes the
universal properties of the eigenvalues in the ergodic phase which depend
only on the system's symmetry while independent of microscopic details.
Specifically, the Gaussian orthogonal/unitary ensemble (GOE/GUE) describes
systems with/without time reversal symmetry, and Gaussian symplectic
ensemble (GSE) stands for time-reversal invariant systems with broken spin
rotational invariance. On the contrary, the eigenvalues in MBL phase are
independent of each other and belongs to the Poisson ensemble.

Compared to the properties of each phase, the evolution between them is much
less understood, especially on the ergodic side. There are two issues of
particular interest. First one concerns the energy scale called Thouless
energy $E_{\text{Th}}$, defined through the Thouless time $t_{\text{Th}%
}=\hbar /E_{\text{Th}}$, which measures the average time a particle takes to
diffuse over the system. Therefore, eigenvalue fluctuations below $E_{\text{%
Th}}$ are well characterize by RMT and hence universal, while those above $%
E_{\text{Th}}$ are model dependent. This energy scale $E_{\text{Th}}$ is
difficult to observe with local spectral statistics such as level spacing or
spacing ratio distributions, but can be probed by long-range spectral
measures such as the number variance\cite%
{Altshuler1986,Berkovits20,Berkovits21} or spectral form factor\cite%
{Ludwig18,Chan18,SierantPRL,Vidmar}. The second issue regards the Griffiths
regime near the transition region\cite{G1,G2,G3,G4,G5}, where an
inhomogeneous mixture of locally thermal and localized regions co-exist and
result in anomalously slow dynamics and multifractality of eigenstates.
Unlike the Thouless energy, study of Griffiths effect is normally based on
eigenfunctions.

In this work, we employ the singular-value-decomposition (SVD) to study the
eigenvalue spectra of random spin systems with MBL transition, and show that
both the Thouless energy and Griffiths regime can be revealed through the
scaling of singular values. The advantage of this method are two-folded. For
the Thouless energy, SVD requires no unfolding procedure, which is necessary
for studying number variance and spectral form factor, and therefore avoids
the potential ambiguity raised by concrete unfolding strategy\cite{Gomez2002}%
. Very recently, Berkovits employed SVD to the study of Thouless energy in
Anderson model\cite{Berkovits21} and the identification of non-ergodic
extended phase in the Rosenzweig-Porter model\cite{Berkovits20,Kravtsov},
and now we bring it to the many-body regime. For the Griffiths effect, the
inputs of SVD are the eigenvalue spectra, which are numerically much easier
to obtain than the eigenfunctions.

The underlying mechanism behind SVD is to view the eigenvalue spectrum of a
complex quantum system as a time-series\cite%
{TimeSeries2002,Vargas1,Vargas2,Vargas3,Garcia2006,Faleiro,Molina}, and by
performing SVD to the ``sample matrix'' (see Eq.~(\ref{equ:X}) in Sec.\ref%
{sec2}) we are able to distinguish the trend and fluctuation modes therein.
This technique is in essence identical to the unsupervised machine learning
algorithm called principal component analysis (PCA), which has also found
various applications in condensed matter physics\cite{PCA1,PCA2,PCA3,PCA4}.
However, while PCA deals with several components with largest weights, we
shall see the universal information of MBL system is encoded in the
intermediate components with lower weights.

This paper is organized as follows. In Sec. \ref{sec2} we introduce the SVD
method, and show the scree plot (singular values as a function of level
index) for the MBL phase reflects a clear integrable behavior, while that
for the ergodic phase breaks into two branches -- a universal high-order
part that belongs to RMT and a non-universal part, with the starting point
of the former identifying the Thouless energy $E_{\text{Th}}$. In Sec. \ref%
{sec3} we dig into the structures of the principal components, and show the
higher-order ones are close to the Fourier modes of the eigenvalue spectrum,
and explain why the higher-order part of scree plot is universal. In Sec. %
\ref{sec4} we demonstrate the scree plot only reflects the exponential part
of the level spacing distribution while being insensitive to the level
repulsion, or equivalently the system's symmetry. Consequently, as detailed
in Sec. \ref{sec5}, it gives an underestimation on the ergodic-MBL
transition point, which suggests a non-ergodic behavior that may be
attributed to the Griffiths regime. Discussion and conclusion come in Sec. %
\ref{sec6}.

\section{SVD on Eigenvalue spectra}

\label{sec2}

\bigskip In this work we consider the paradigmatic spin model with
ergodic-MBL transition, that is, the anti-ferromagnetic Heisenberg model
with random external fields\cite{Alet}, the Hamiltonian reads%
\begin{equation}
H=\sum_{i=1}^{L}\mathbf{S}_{i}\cdot \mathbf{S}_{i+1}+\sum_{i=1}^{L}%
\sum_{\tau =x,y,z}h_{\tau }\varepsilon _{i}^{\tau }S_{i}^{\tau }\text{,}
\label{equ:H}
\end{equation}%
where the coupling strength is set to be $1$, and $\varepsilon _{i}^{\tau }$%
s are random variables within range $\left[ -1,1\right] $. We first consider
the orthogonal case with time-reversal symmetry, that is, $h_{x}=h_{z}=h\neq
0$ and $h_{y}=0$, where an ergodic-MBL transition happens at $h_{c}\simeq 3$%
\cite{Regnault16,Regnault162}. Using exact diagonalization, we generate $%
N=1000$ samples of eigenvalue spectra at various randomness $h$ in an $L=13$
system, with the Hilbert space dimension being $2^{13}=8192$. For each
spectrum, we take out the middle $P=1000$ eigenvalues, and arrange them into
an $N\times P$ matrix $X$,
\begin{equation}
X=\left(
\begin{array}{cccccc}
E_{1}^{\left( 1\right) } & E_{2}^{\left( 1\right) } & . & . & . &
E_{P}^{\left( 1\right) } \\
E_{1}^{\left( 2\right) } & E_{2}^{\left( 2\right) } & . & . & . &
E_{P}^{\left( 2\right) } \\
. & . & . &  &  & . \\
. & . &  & . &  & . \\
. & . &  &  & . & . \\
E_{1}^{\left( N\right) } & E_{2}^{\left( N\right) } & . & . & . &
E_{P}^{\left( N\right) }%
\end{array}%
\right) \text{,}  \label{equ:X}
\end{equation}%
where $E_{i}^{\left( j\right) }$ stands for the $i$-th eigenvalue in the $j$%
-th sample. For clarity, we shall call $X$ the ``sample matrix''. We then
perform SVD on $X$, which equals to re-express $X$ as%
\begin{equation}
X=U\Lambda W\equiv \sum_{k}\sigma _{k}X^{\left( k\right) }\text{, }%
X_{ij}^{\left( k\right) }=U_{ik}W_{kj}\text{,}
\end{equation}%
where $\Lambda $ is an $N\times P$ matrix whose non-zero elements $\Lambda
_{i,i}\equiv \sigma _{i}$ are the ordered singular values $\sigma _{1}\geq
\sigma _{2}\geq ...\geq \sigma _{r}$ with $r\leq \min [N,P]=$ Rank$[X]$.
This technique is equivalent to the machine learning algorithm called
principal component analysis (PCA), the spirit of which is to view the
eigenvalue spectrum as a multi-dimensional data, and by SVD we decompose it
into orthonormal modes $W_{k}$ -- the $k$-th row of the $P\times P$ matrix $W
$ -- with weight $\sigma _{k}$. The $W_{k}$ is called principal component in
the terminology of PCA, and encodes one feature (character) of the
eigenvalue spectrum, whose physical meaning is read by evaluating its
behavior $W_{k}[i]$, where $i$ denotes the energy level index. Given the decreasing
tendency of $\sigma _{k}$, we can approximate the original sample matrix $X$
by $\widetilde{X}=\sum_{k=1}^{m}\sigma _{k}X^{\left( k\right) }$ with some
value $m$, hence achieving the purpose of dimension reduction. In most
applications of PCA, we only keep several dominant components with largest
weights. However, as we shall see, the universal information of the random
spin system is encoded in the intermediate components with lower weights.

Another interpretation of SVD is to treat $X$ as a multivariate time series,
with $\lambda _{k}\equiv \sigma _{k}^{2}$ being the eigenvalue of the
covariance matrix $X^{T}X$. It has been demonstrated \cite%
{Vargas1,Berkovits20,Berkovits21} that $\lambda _{k}$ with large $k$ follows
the power-law behavior%
\begin{equation}
\lambda _{k}\sim \frac{1}{k^{\alpha }}  \label{equ:power}
\end{equation}%
with $\alpha =1\left( 2\right) $ in the chaotic (integrable) system,
corresponding to the ergodic (MBL) phase respectively.

\begin{figure}[t]
\centering\includegraphics[width=\columnwidth]{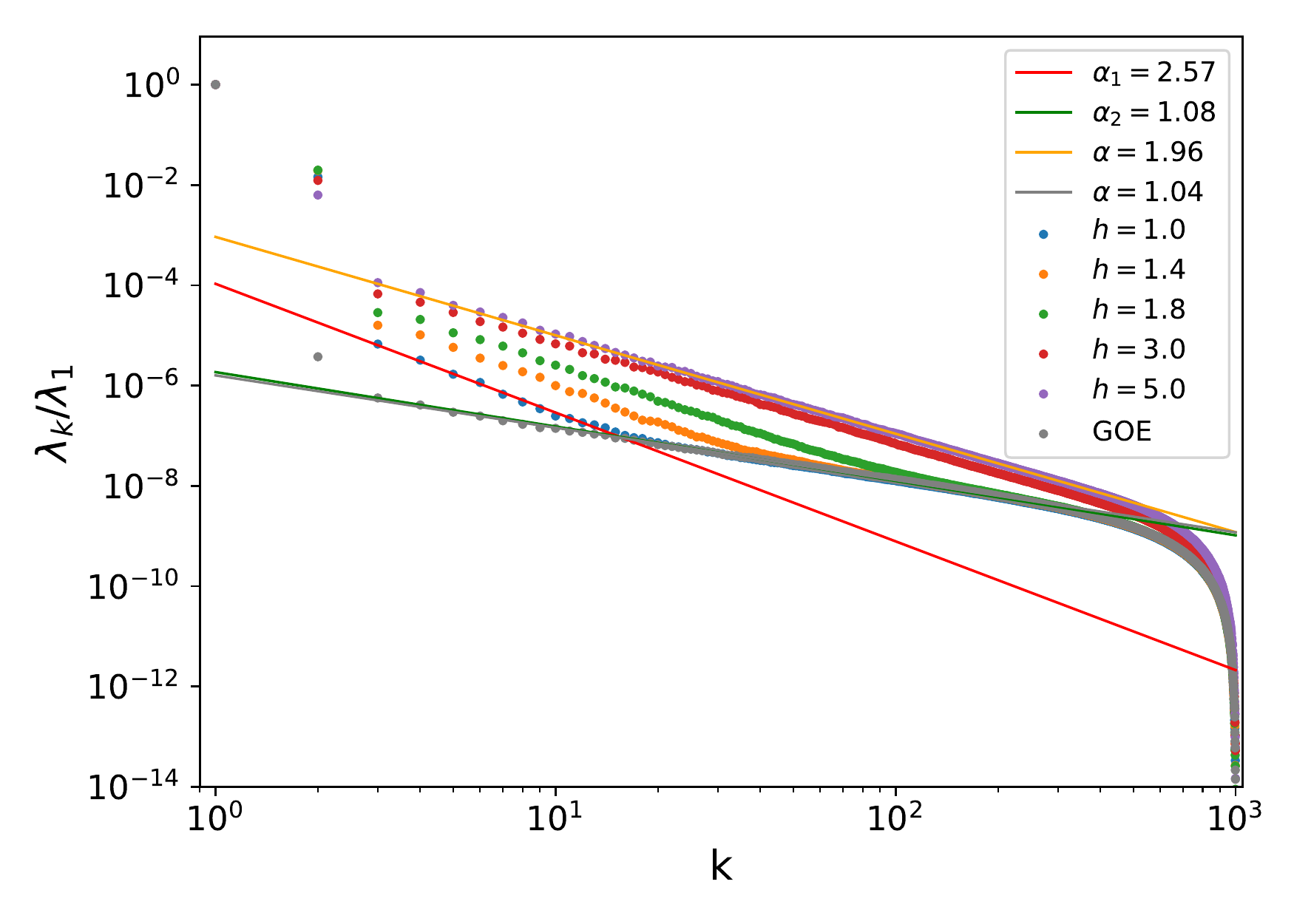}
\caption{Scree plots of $\left\{\lambda_k/\lambda_1\right\}$ at various randomness strengths.
Despite the two dominant modes, $\lambda_k$ in the MBL phase ($h=5$) follows
a power-law $k^{-\alpha}$ with $\alpha\simeq2$ reflecting the integrable behavior.
Two-branch structures appear for $\lambda_k$ in the ergodic phase ($h<3$),
separating the model-dependent lower part and the universal higher
part that belongs to RMT, the starting point of the higher part provides an
estimation of the Thouless energy.} \label{fig:demo}
\end{figure}

As a demonstration, we generate the sample matrix $X$ for Eq.(\ref{equ:H})
with $L=13$ at various randomness strengths, and plot the resulting $\lambda
_{k}$ after SVD with respect to index $k$ (the so-called scree plot) in a
log-log scale in Fig.~\ref{fig:demo}. Without loss of generality, we divide
each $\lambda _{k}$ by the largest value $\lambda _{1}$. For comparison, we
also generate eigenvalue spectra from modeling GOE matrices (orthogonal
matrices with random elements drawn from standard Gaussian distribution $%
N\left( 0,1\right) $) of the same size, the resulting $\lambda _{k}/\lambda
_{1}$ appear as the grey dots therein. Clearly, in all cases, the first two
modes are orders of magnitudes larger than the rest, which indicates the
existence of two overwhelming features in the eigenvalue spectrum. For
reasons detailed in next section, we identify them to be the mean energy $%
\langle E\rangle $ and mean level spacing $\langle s\rangle $. While for
large modes with $k>300$, the weights drops very rapidly, indicating they
contribute little to the properties of the eigenvalue spectrum. The
universal information is thus expected to be encoded in the intermediate
modes with $2<k<300$.

\bigskip More specifically, for case in the MBL\ phase ($h=5$), the scree
plots with $3\leq k\leq 200$ fits well into the power-law behavior Eq.(\ref%
{equ:power}) with $\alpha =1.962$, very close to the value $2$ for
integrable system as expected. While for data in the ergodic phase ($h=1$), $%
\lambda _{k}$ divides into two parts: the higher-part $30\leq k\leq 200$
with power-law exponent $\alpha _{2}\simeq 1.085$ almost coincides with that
of modelling GOE, reflecting a clear chaotic behavior, and this part is
identified to be universal below Thouless energy and related to RMT; while
the lower-part $3\leq k\leq 15$ follows a super-Poissonian power-law
behavior with $\alpha _{1}\simeq 2.567$, and is therefore identified to be
the model-dependent part beyond Thouless energy, the more detailed analysis
will be provided in Sec.III. Interestingly, this value of $\alpha _{1}$
happens to be close to the one in the metallic phase of a 3D Anderson model%
\cite{Berkovits21}, which may suggest a subtle correspondence between these
two models.

From Fig.~\ref{fig:demo} we can see such a two-branch scree plot is general
for data in ergodic phase, and quantitatively we identify the Thouless
energy $E_{\text{Th}}$ through the starting point of the chaotic behavior $%
\lambda _{k}\sim k^{-1}$, which is $k_{\text{Th}}\sim 30$ for the case $h=1$%
.\ As the randomness grows, the $k_{\text{Th}}$ increases -- which indicates
$E_{\text{Th}}$ decreases -- and the two branches gradually get mixed. At
the transition point $h\simeq 3$, the scree plot becomes almost identical to
that of the MBL phase, suggesting the saturation of $\alpha _{2}$ may be an
indicator for the MBL transition. However, as discussed in detail in Sec.~%
\ref{sec5}, this may not provide an accurate estimation.

Up to now, we have shown the scree plots provides a transparent way to
reveal the Thouless energy $E_{\text{Th}}$. However, since there's artifact
in determining where the chaotic behavior $\lambda _{k}\sim k^{-1}$ begins
to hold, this method is less accurate in quantitatively determining $E_{\text{%
Th}} $ than more standard probes like spectral form factor. Meanwhile, the
numerics clearly indicate that $E_{\text{Th}}$ decreases with randomness
strength, which is qualitatively consistent with the results in Ref.~[%
\onlinecite{Vidmar}]. Moreover, as stated in Sec.I, the central advantage of
SVD is that it requires no unfolding procedure. To further pursuit the
physics behind the scree plot, we now turn to the analysis of principal
components $W_{k}$.

\section{Principal Component Analysis}

\label{sec3}

The $k$-th principal component of the eigenvalue spectrum is represented by the $%
k$-th row of the matrix $W$, we therefore evaluate the behavior of $W_{k}$ with
respect to the energy level index $i$ to read out the physics. The results
in this section are based on the data from $h=1$, and we have checked they
hold in other cases as well.

\begin{figure}[t]
\centering
\includegraphics[width=\columnwidth]{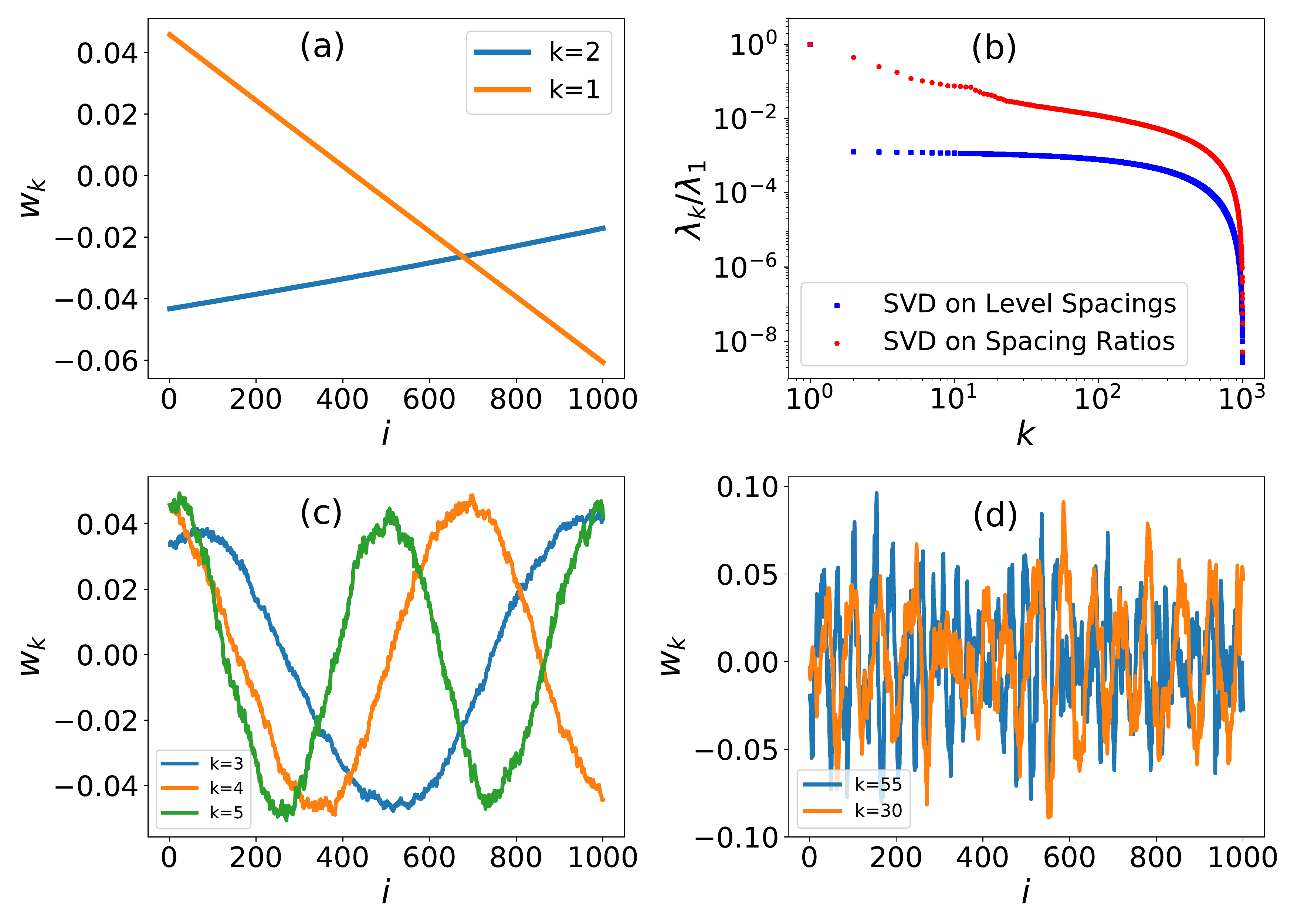}
\caption{(a) Behaviors of the two dominant components $W_{1/2}$, the linear behaviors
indicate they correspond to two non-fluctuating features of the eigenvalue spectrum, which
are suspected to be the mean energy $\langle E\rangle$ and level spacing $\langle s\rangle$.
(b) Scree plots of $\lambda_k$ after performing SVD on the sample matrix comprised of
level spacings and spacing ratios, where the ratio of first two weights $\lambda
_{1}/\lambda _{2}=789.14\left( 2.25\right) $ in respective cases. (c),(d) Behaviors of typical higher components $W_k$, the quasi-sinusoidal
behaviors indicate they are close to the $k$-th Fourier mode of the eigenvalues,
and the frequency increases with $k$.
} \label{fig:PCA}
\end{figure}

We have seen in Fig.~\ref{fig:demo} that the scree plot is dominated by two
largest weights $\lambda _{1}$ and $\lambda _{2}$, which means the
eigenvalue spectrum contains two dominant features that are overwhelming over
other features. We therefore draw the behaviors of two dominant modes $W_{1}$
and $W_{2}$ in Fig.~\ref{fig:PCA}(a), it's clear that both of them are
linearly dependent with level index $i$, suggesting they correspond to two
non-fluctuating features of the eigenvalue spectrum. We postulate these two
dominant features to be the mean energy $\langle E\rangle $ and the mean
level spacing $\langle s\rangle $, both of which are non-universal. To
support this conjecture, we first change the input data from the eigenvalues
$\left\{ E_{i}\right\} $ to the level spacings $\left\{
s_{i}=E_{i+1}-E_{i}\right\} $. With this input, the information of mean
energy $\langle E\rangle $ is lost, leaving $\langle s\rangle $ as the only
dominant feature. Therefore when applying SVD to this new sample matrix $%
X^{\prime }$, we should observe only one dominant weight $\lambda _{1}$.
This is verified in Fig.~\ref{fig:PCA}(b), where $\lambda _{1}/\lambda
_{2}=789.14$ is observed. We can further change the input to be the spacing
ratios $\left\{ s_{i+1}/s_{i}\right\} $, where the information about $%
\langle E\rangle $ and $\langle s\rangle $ are both lost, and the resulting
scree plot is expected to contain no dominant weights. This is also verified
in Fig.~\ref{fig:PCA}(b), where we see $\lambda _{1}$ is the same order with
$\lambda _{2}$ (the precise value is $\lambda _{1}/\lambda _{2}=2.25$),
which confirms our conjecture.

Strictly speaking, the mean energy $\langle E\rangle $ and level spacing $%
\langle s\rangle $ are non-universal for different reasons. The $\langle
E\rangle $ stands for the global energy scale of the system and therefore
depends on the model's details. While $\langle s\rangle $ is the value that
can be artificially assigned when counting level statistics, which stems
from the mathematical degree of freedom in the joint probability
distribution function of the eigenvalues. To be specific, recall the eigenvalue
distribution function in the standard WD class is\cite{Mehta,Haake}%
\begin{equation}
P\left( \left\{ E_{i}\right\} \right) =A_{1}\prod_{i<j}\left\vert
E_{i}-E_{j}\right\vert ^{\beta }e^{-A_{2}\sum_{i}E_{i}^{2}};\beta =1,2,4%
\text{,}  \label{equ:WD}
\end{equation}%
where $A_{1}$ and $A_{2}$ are two parameters constrained by single
normalization condition $\int P\left( \left\{ E_{i}\right\} \right)
\prod_{i}dE_{i}=1$. From $P\left( \left\{ E_{i}\right\} \right) $ we can
derive the celebrated WD distribution $P\left( s\right) =A_{1}s^{\beta
}e^{-A_{2}s^{2}/2}$ using Wigner surmise\cite{Mehta,Haake}.\ The single
normalization condition $\int P\left( s\right) ds=1$ is not sufficient to
determine two parameters $A_{1},A_{2}$, which gives us the freedom to choose
$\langle s\rangle $. In most practical studies, we take $\langle s\rangle $
to be $1$.

Having understood the physics of the two dominant components, we now return
to analysis of the components $W_{k}$ $\left( k\geq 3\right) $ of the
original sample matrix $X$, several of which are plotted in Fig.~\ref%
{fig:PCA}(c),(d). We observe clear sinusoidal behaviors of $W_{k}$ with
respect to level index $i$, with the frequency increasing with $k$.
Specifically, we have $W_{k}[i]\sim \cos \left( \frac{\left( k-1\right) \pi
}{N}i+\varphi _{k}\right) $ with $N=1000$ being the number of eigenvalues,
which indicates the component $W_k$ is close to the $k$-th Fourier mode of the
eigenvalue spectrum. Therefore, for smaller $k$, $W_{k}$ reflects the level
fluctuations on larger energy scale. When $k$ is not so large, this energy
scale goes beyond Thouless energy, so its behavior is non-universal; while
when $k$ exceeds certain threshold $k_{\text{Th}}$, this energy scale falls
within Thouless energy. This threshold $k_{\text{Th}}$ divides the $%
\lambda_k $ into two parts: the non-universal part with $k<k_{\text{Th}}$
and universal part with $k>k_{\text{Th}}$. This is the origin of the
two-branch structure of the scree plot in the ergodic phase.

\begin{figure}[t]
\centering
\includegraphics[width=\columnwidth]{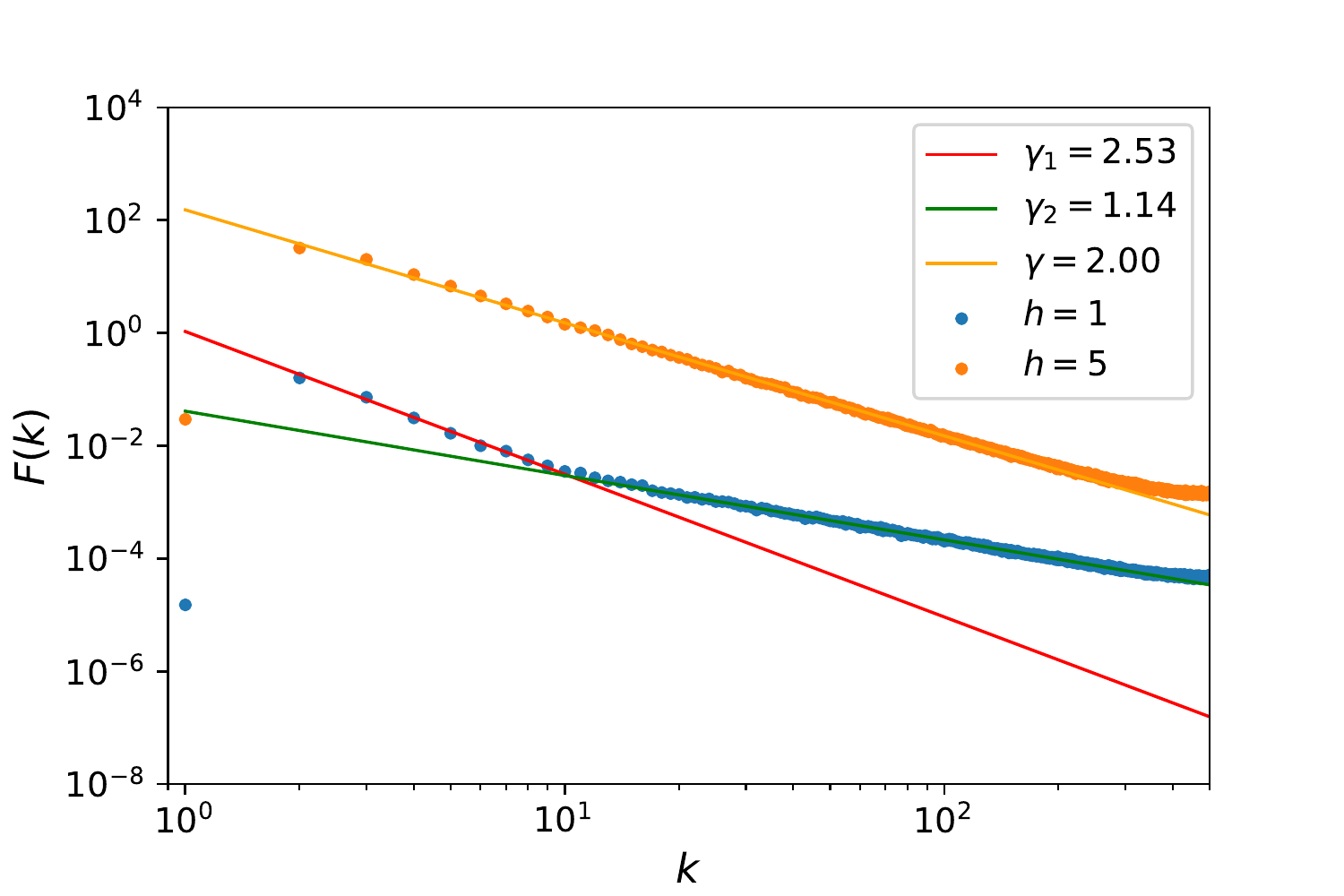}
\caption{Power spectrum function $F(k)$ for ergodic ($h=1$) and MBL ($h=5$) phases, in the former
a clear two-branch structure appears just like in the scree plot of $\lambda_k$. The power-law
exponents $\gamma\simeq\alpha$ are observed in all cases.} \label%
{fig:power}
\end{figure}

\bigskip The local fluctuations of the eigenvalue spectra can be further
revealed by considering the power spectrum function, whose definition is
\begin{equation}
F_{m}\left( k\right) =\left\vert \frac{1}{r}\sum_{n=1}^{r}\left[ \left(
\sum_{p=3}^{r}\sigma _{p}X_{mn}^{\left( p\right) }\right) \exp \left( -\frac{%
2\pi ink}{r}\right) \right] \right\vert ^{2}\text{,}  \label{equ:Fk}
\end{equation}%
where $\sum_{p=3}^{r}\sigma _{p}X_{mn}^{\left( p\right) }$ are the
eigenvalue spectra after ``global
unfolding''\cite{Vargas1,Vargas2,Vargas3}. By averaging
over the ensemble we get the power spectrum function $F\left( k\right) =%
\frac{1}{N}\sum_{m=1}^{N}F_{m}\left( k\right) $, which was shown to behave
differently in integrable and chaotic systems\cite%
{TimeSeries2002,Vargas1,Berkovits21,E1,E2}. To this end, we calculate $F\left(
k\right) $ in the $h=1\left( 5\right) $ case standing for the ergodic (MBL)
phase, the results are shown in Fig.~\ref{fig:power}. As we can see, $%
F\left( k\right) $ behaves totally similar to $\lambda _{k}$: in the ergodic
phase $F\left( k\right) $ divides into two branches that both follow $%
1/k^{\gamma }$, specifically, the lower modes satisfy $\gamma _{1}\simeq
2.5\simeq \alpha _{1}$ and higher modes satisfy $\gamma _{2}\simeq 1\simeq
\alpha _{2}$; while for the MBL case, a single curve with $\gamma \simeq
2\simeq \alpha $ appears. This coincidence has also appeared in the
non-interacting Anderson model\cite{Berkovits21}, which indicates that $%
F\left( k\right) $ and $\lambda _{k}$ essentially contain identical physical
information. Based on the PCA results above, we can provide a qualitative
explanation as follows.

We have verified that $W_{k}$ is close to the Fourier modes of eigenvalue
spectrum for $k\geq 3$, hence the power-law behavior $\lambda _{k}\sim k^{-\alpha }$
essentially indicates a decreasing trend of the eigenvalues' Fourier weights.
On the other hand, the definition of $F\left( k\right) $ in Eq.~(\ref{equ:Fk}%
) drops the first two dominant terms $X^{\left( 1/2\right) }$, which
stand for the mean energy $\langle E\rangle $ and level spacing $\langle
s\rangle $ that are both non-fluctuating. Therefore, the fluctuating
behaviors of original eigenvalue spectra and the one after ``global
unfolding'' (that is, $\sum_{p=3}^{r}\sigma _{p}X_{mn}^{\left( p\right) }$)
should be the same. Consequently, the scaling behaviors of $\lambda _{k}$
and $F\left( k\right) $ are expected to be identical.

A technical issue worth mentioning is that the choice of power spectrum in
Eq.~(\ref{equ:Fk}) is not unique. Another routine is to construct the
following time-series\cite{TimeSeries2002,Garcia2006,Faleiro,Molina}
\begin{equation}
\delta _{n}=\sum_{i=1}^{n}\left( s_{i}-\langle s\rangle \right)
=s_{n}-n\langle s\rangle \text{,}
\end{equation}%
where $s_{n}$ is the $n$-th order level spacing. It is clear that $\langle
\delta _{n}\rangle $ gets rid of the non-universal information about $%
\langle E\rangle $ and $\langle s\rangle $, in a similar way to $%
\sum_{k=3}^{r}\sigma _{k}X_{ij}^{\left( k\right) }$ in Eq.~(\ref{equ:Fk}),
hence its Fourier weight $S\left( k\right) $ show totally similar behaviors
to $F(k)$ in Fig.~\ref{fig:power}. Despite these qualitative estimations, we
also note the analytical derivation for $S(k)$ is recently given in Ref.~[%
\onlinecite{Kanzieper1,Kanzieper2,Kanzieper3}].

\section{Relating $\protect\alpha$ To Level Spacing Distribution}

\label{sec4}

Up to now, we have demonstrated that the power-law exponent $\alpha$
(which is $\alpha_2$ in the ergodic phase) for $\lambda _{k}$ with
large $k$ is universal that related to RMT. However, the
power-law exponent $\alpha $ is only a single number, it's still
questionable whether it can distinguish different random matrix ensembles
which may contain multiple parameters. For example, the general form of
level spacing distribution $P\left( s\right) $ contains two parts: the
polynomial part that reflects the level repulsion, and an exponential part
that reflects large $s$ decaying. As for the random spin system, a
widely-used two-parameter spacing distribution for the ergodic-MBL
transition is proposed by Serbyn and Moore\cite{Serbyn}, i.e.%
\begin{equation}
P\left( \beta _{1},\beta _{2},s\right) =C_{1}s^{\beta _{1}}\exp \left(
-C_{2}s^{2-\beta _{2}}\right) \text{.}  \label{equ:dist}
\end{equation}%
For case deep in the ergodic phase, we have $\beta _{1}=1,\beta _{2}=0$
standing for the GOE distribution; while for case deep in the MBL phase, $%
\beta _{1}=0,\beta _{2}=1$ for the Poisson ensemble. We conjecture that the
power exponent $\alpha $ of $\lambda _{k}$ corresponds \emph{only} to the
parameter $\beta _{2}$ in the exponential part, while being insensitive to the
level repulsion $\beta _{1}$. If this conjecture is correct, two deductions
are immediate: (i) the scree plot with $\alpha =2$ does not necessarily
correspond to the Poisson ensemble, but also to a number of intermediate
ensembles whose spacing distribution decays as $e^{-Cs}$; (ii) the scree
plot with $\alpha =1$ in ergodic phase is insensitive to the level
repulsion, or equivalently the symmetry of the system.

\begin{figure*}[t]
\centering\includegraphics[width=2\columnwidth]{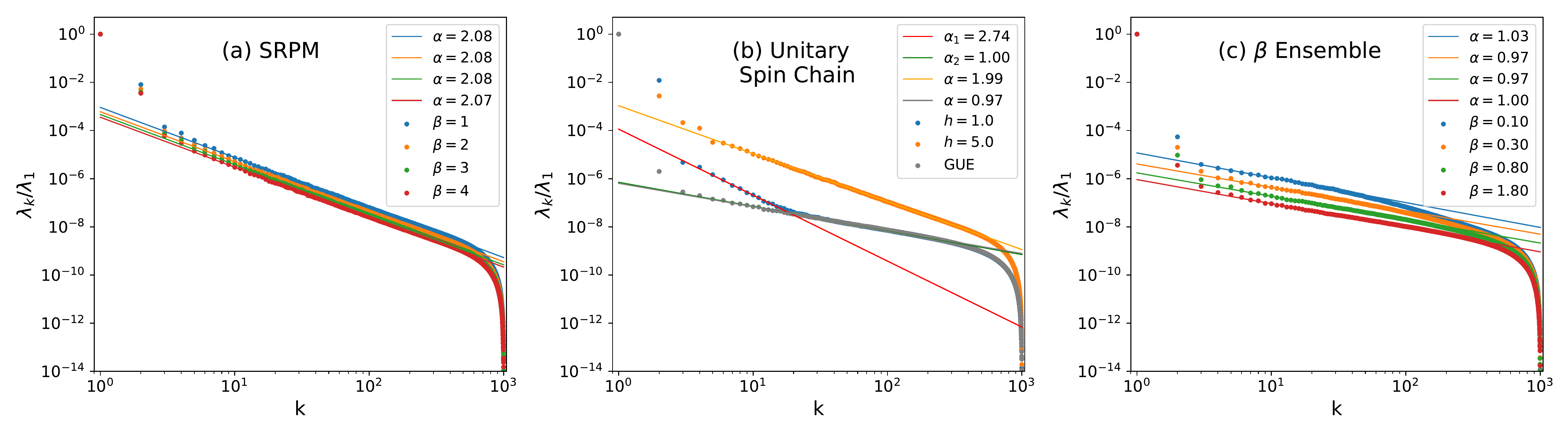}
\caption{(a) $\left\{\protect\lambda_k/\protect\lambda_1\right\}$ of SRPM
with different level repulsion $\protect\beta$, the integrable behavior
 $\lambda_k\sim k^{-2}$ appears in all cases. (b) $\left\{\protect\lambda_k/\protect%
\lambda_1\right\}$ for the unitary spin system, the behaviors of both the
ergodic ($h=1$) and MBL ($h=5$) phases are totally similar to those in the
orthogonal system in Fig.~\ref{fig:demo}, indicating the scree plot is insensitive to the system's
symmetry. (c) $\left\{\protect\lambda_k/\protect\lambda_1\right\}$ for the
Gaussian $\protect\beta$ ensemble with several non-standard values of $%
\protect\beta$, chaotic behavior $\lambda_k\sim k^{-1}$ appears in all
cases.}
\label{fig:SRPM}
\end{figure*}

To verify (i), we consider the random matrix model called short-range plasma
model (SRPM)\cite{SRPM}, which describes the eigenvalues as an ensemble of
one-dimensional particles with only nearest neighboring logarithmic
interactions. SRPM holds the semi-Poisson distribution which is widely
accepted as the critical distribution at the MBL transition point. It is
shown\cite{SRPM} that the large $s$ behavior of SRPM decays as $s^{\beta
}e^{-\left( \beta +1\right) s}$ with $\beta $ being the Dyson index
controlling the strength of level repulsion. Thus, if the deduction (i)
holds, we should observe identical power-law exponents $\alpha $ for cases
with different $\beta $. To effectively obtain the eigenvalue spectrum of
the SRPM, we make use of an elegant correspondence between SRPM and the
Poisson ensemble, that is, the spectrum comprised of every $\left( \beta
+1\right) $-th eigenvalue from the Poisson ensemble is identical to the
eigenvalue spectrum of SRPM with index $\beta $\cite{Daisy}. Therefore, we
can easily obtain the sample matrix of SRPM to perform SVD, and the
resulting scree plots for $\beta =\{1,2,3,4\}$ are displayed in Fig.~\ref%
{fig:SRPM}(a). As expected, the scree plots are totally
similar, and the power-law exponent $\alpha \simeq 2$ appears in all cases,
which verifies (i).

For deduction (ii), we first consider a unitary spin system that breaks
time-reversal symmetry, i.e. the case with $h_{x}=h_{y}=h_{z}=h\neq 0$ in
Eq.~(\ref{equ:H}). This model undergoes an ergodic-MBL transition at around $%
h_{c}\simeq 2.5$\cite{Regnault16,Regnault162}, with the level spacing
distribution evolving from GUE to Poisson. We likewise take $h=1$ and $h=5$
to represent the ergodic and MBL phase respectively. With the same size of
samples, we obtain the resulting scree plots shown in Fig.~\ref{fig:SRPM}%
(b), where we also draw $\left\{ \lambda _{k}/\lambda _{1}\right\} $ from a
modeling GUE for comparison. As can be seen, $\lambda _{k}\sim 1/k$ appears
in both the higher part of ergodic phase and modeling GUE, which are totally
similar to the cases in Fig.~\ref{fig:demo}; and $\lambda _{k}\sim 1/k^{2}$
appears for data in MBL phase as expected. We further consider
the Gaussian $\beta $ ensemble which holds the same form of eigenvalue
distribution as WD classes (Eq.~(\ref{equ:WD})) but with $\beta $ different from
$\{1,2,4\}$. The eigenvalue spectra of general Gaussian $\beta$ ensemble can
be efficiently generated by diagonalizing the following tridiagonal matrix\cite{Beta}%
\begin{equation}
M_{\beta }=\frac{1}{\sqrt{2}}\left(
\begin{array}{ccccc}
x_{1} & y_{1} &  &  &  \\
y_{1} & x_{2} & y_{2} &  &  \\
&
\begin{array}{ccc}
\text{.} &  &  \\
& \text{.} &  \\
&  & \text{.}%
\end{array}
&
\begin{array}{ccc}
\text{.} &  &  \\
& \text{.} &  \\
&  & \text{.}%
\end{array}
&
\begin{array}{ccc}
\text{.} &  &  \\
& \text{.} &  \\
&  & \text{.}%
\end{array}
&  \\
&  & y_{N-2} & x_{N-1} & y_{N-1} \\
&  &  & y_{N-1} & x_{N}%
\end{array}%
\right)  \label{equ:Beta}
\end{equation}%
where the diagonals $x_{i}\,$($i=1,2,...,N$) follow the normal distribution $%
\mathit{N}\left( 0,2\right) $, and $y_{k}$ ($k=1,2,...,N-1$) follows the $%
\chi $ distribution with parameter $\left( N-k\right) \beta $. Without loss
of generality, we select several non-standard values of $\beta $, and
generate $1000$ samples of eigenvalue spectrums with the matrix dimension $%
N=4000$ to construct the sample matrix $X$. The resulting scree plots are
displayed in Fig.~\ref{fig:SRPM}(c), we see the chaotic behavior $\lambda
_{k}\sim k^{-1}$ appears in all cases as expected. Thus, combining Fig.~\ref%
{fig:SRPM}(b) and (c) we confirm deduction (ii).

\section{Evolution of $\protect\alpha_2$ During Ergodic-MBL Transition}

\label{sec5}

Given the universal information of the system is encoded in the power
exponent $\alpha _{2}$ of the higher part of the scree plot, it's a natural
idea to employ it to detect the ergodic-MBL transition, without referring to
the study of eigenfunctions. We expect the exponent $\alpha _{2}$ to evolve
from $1$ in ergodic phase to $2$ in MBL phase. Specifically, we consider the
orthogonal spin model Eq.~(\ref{equ:H}) with length $L=13$ in the randomness
range $h\in \left[ 1,5\right] $ with interval $\delta h=0.2$. We generate $%
1000$ samples of eigenvalue spectra at each $h$, and select $1000$
eigenvalues in the middle to construct the sample matrix $X$ and hence
obtain $\lambda _{k}$. The exponent $\alpha _{2}$ is determined by fitting $%
\lambda _{k}\sim k^{-\alpha _{2}}$ for $50<k<250$ in all cases.

\begin{figure}[t]
\centering
\includegraphics[width=\columnwidth]{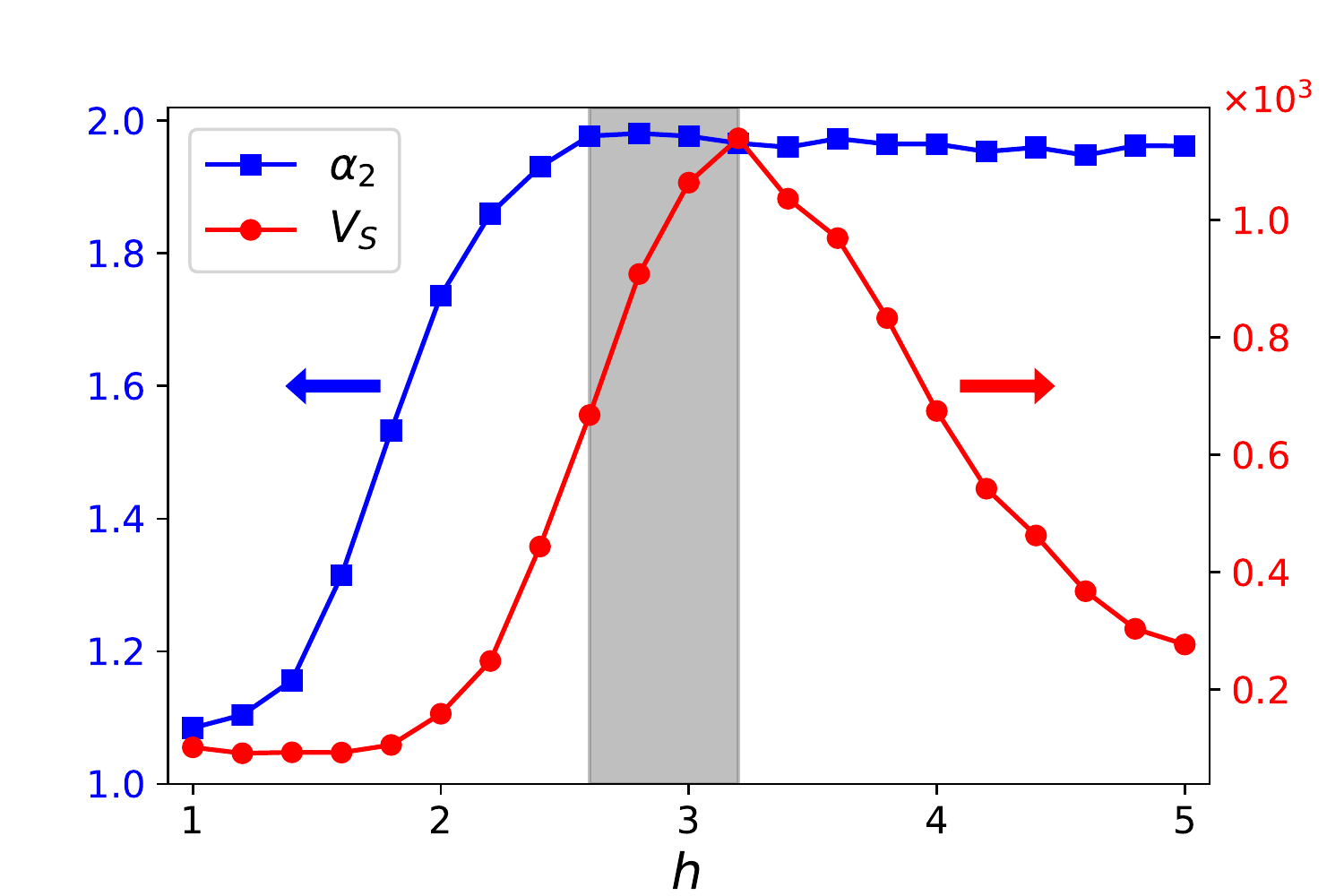}
\caption{Evolutions of $V_S$ and the power-law exponent $\alpha_2$
with respect to the randomness strength $h$.
The grey shaded area represents a non-ergodic region
that may be attributed to the Griffiths regime.} \label{fig:evolution}
\end{figure}

For comparison, we employ another eigenvalue-based quantity to detect the
MBL transition, that is, the inter-sample randomness, whose definition goes
as follows. First, we adopt a variant definition of spacing ratio, which is%
\cite{Oganesyan}%
\begin{equation}
t_{i}=\frac{\min \left\{ s_{i+1},s_{i}\right\} }{\max \left\{
s_{i+1},s_{i}\right\} }\text{,}
\end{equation}%
where $s_{i}=E_{i+1}-E_{i}$ is the $i$-th level spacing. The mean value $%
\overline{t}$ in different ensembles has been calculated to be\cite{Atas} $%
\overline{t}_{\text{GOE}}=0.536$ and $\overline{t}_{\text{Poisson}}=0.386$.
The calculation of $\overline{t}$ contains two steps: first we calculate the
mean spacing ratio $t_{S}=\langle t_{i}\rangle _{\text{samp}}$ in one
sample, then we average $t_{S}$ over an ensemble of samples to get $%
\overline{t}=\langle t_{S}\rangle _{\text{en}}$. The two steps generates two
types of variance, the first one is $V_{S}=\langle t_{S}^{2}-\overline{t}%
^{2}\rangle _{\text{en}}$, i.e. the variance of sample-averaged spacing
ratio over ensemble, which measures the inter-sample randomness; another one
is $V_{I}=\langle v_{I}\rangle _{\text{en}}$ where $v_{I}=\langle
t_{i}^{2}-t_{S}^{2}\rangle _{\text{samp}}$, measuring the intrinsic
intra-sample randomness. In an MBL system driven by random disorder such as
Eq.~(\ref{equ:H}), the distribution of $t_{S}$ near the transition region
will deviate from a Gaussian function -- a manifestation of Griffiths regime
-- which results in a peak of $V_{S}$ at the transition point\cite%
{RD,Sierant19,Rao21}. Therefore, we can compute the evolution of $V_{S}$ to
locate the transition.

The evolution of $V_{S}$ and $\alpha _{2}$ are drawn collectively in Fig.~%
\ref{fig:evolution}. The peak of $V_{S}$ identifies the transition point to
be $h_{c}\simeq 3.2$, which is very close to the widely accepted value $%
h_{c}\simeq 3$\cite{Regnault16,Regnault162}, while $\alpha _{2}$ saturates
to $2$ at roughly $h=2.6$, smaller than the transition value. This clearly
indicates a non-ergodic region in the ergodic phase.

\bigskip One explanation for this result is the existence of Griffiths
regime close to the transition point that contains locally localized small
regions. To be specific, the evolution of the spacing distribution in Eq.~(%
\ref{equ:dist}) during the ergodic-MBL transition was discussed by Serbyn and Moore\cite%
{Serbyn}, where they showed that Griffiths effects affect the large $s$
decaying of $P(s)$ more than the small $s$ behavior. Consequently the
parameter $\beta _{2}$ becomes $\beta _{2}\sim 1$ even when the system
is globally in the metallic phase. As demonstrated in previous section, $%
\alpha _{2}$ detects only the power exponent in the exponential part of $%
P(s) $ (that is, $\beta _{2}$), it is not surprising that $\alpha _{2}$
saturates earlier than the transition happens. However, we cannot rule out
other possibilities such as the existence of non-ergodic extended phase\cite%
{Sieber,Khaymovich1,Khaymovich2,Khaymovich3,Altshuler1997} without studying
the distribution of eigenvectors. A possible way to gain further evidence
would be to study MBL system with quasi-periodic potential that is free of
Griffiths regime, it's intriguing to see whether $\alpha_2 $ will
underestimate the transition point in such systems. This, however, goes beyond the
scope of current work.

\section{Conclusion and Discussion}

\label{sec6}

We have employed the singular value decomposition (SVD) to study the
eigenvalue spectra of the random spin systems. By treating the eigenvalue
spectrum as a time-series, we show the
eigenvalue spectrum are dominated by two non-universal features -- the mean
energy $\langle E\rangle $ and level spacing $\langle s\rangle $, while
higher component $W_{k}\left( k\geq 3\right) $ is close to the spectrum's $k$%
-th Fourier mode. Consequently, the scree plot of the singular values $\lambda
_{k}$ in the ergodic phase exhibits two-branch structure: a non-universal (model-dependent)
lower-order part and a universal higher-order part that belong to RMT, and the starting point
of the latter gives an estimation for the Thouless energy. Compared to more
standard probes like number variance or spectral form factor, this
approach requires no-unfolding procedure.

Moreover, we also improve the understanding about the power-law
behavior $\lambda _{k}\sim k^{-\alpha }$. By studying the scree plots of
SRPM, unitary spin chain and non-standard Gaussian $\beta$ ensembles,
we verified the universal exponent $\alpha $ corresponds only to the
parameter $\beta _{2}$ in the exponential part of level spacing distribution
in Eq.~(\ref{equ:dist}) while being insensitive to
the level repulsion parameter $\beta _{1}$, or equivalently the symmetry of
the system. Consequently, $\alpha $ gives an underestimation for the
ergodic-MBL transition point, which suggests a non-ergodic behavior
that may be attributed to Griffiths regime.

The existence of subtle correspondence between the eigenvalues and
eigenfunctions is an intriguing property of complex quantum system\cite%
{Chalker}, and there is certainly more hidden information in the eigenvalues
deserves to be explored. One immediate direction is to construct an
eigenvalue-based time series that is able to read the symmetry of the system
(most probably through the level repulsion). It's also possible to apply SVD
to non-Hermitian MBL systems with complex eigenvalues\cite{Complex1,Complex2}%
, it's exciting to see whether such power-law scree plots appear in these
systems or not.

\section{Acknowledgements.}

The author acknowledges E. Kanzieper for introducing Ref.~[51-53]. This work
is supported by the National Natural Science Foundation of China through
Grant No.11904069.

\end{document}